\documentclass[twocolumn,english,twocolumn,english,aps,prb,floats]{revtex4}
\usepackage[T1]{fontenc}
\usepackage[latin9]{inputenc}
\usepackage{bm}
\usepackage{amsmath}
\usepackage{amssymb}
\usepackage{graphicx}
\usepackage{esint}
\usepackage{adjustbox}
\usepackage{longtable}
\makeatletter
\@ifundefined{textcolor}{}
{%
 \definecolor{BLACK}{gray}{0}
 \definecolor{WHITE}{gray}{1}
 \definecolor{RED}{rgb}{1,0,0}
 \definecolor{GREEN}{rgb}{0,1,0}
 \definecolor{BLUE}{rgb}{0,0,1}
 \definecolor{CYAN}{cmyk}{1,0,0,0}
 \definecolor{MAGENTA}{cmyk}{0,1,0,0}
 \definecolor{YELLOW}{cmyk}{0,0,1,0}
}

\usepackage{babel}

\makeatother

\usepackage{babel}
\begin{document}

\title{Dynamics of the collapse of a ferromagnetic skyrmion in a centrosymmetric lattice}

\author{A. Derras-Chouk, E. M. Chudnovsky, and D. A. Garanin}

\affiliation{Physics Department, Herbert H. Lehman College and Graduate School,
The City University of New York, 250 Bedford Park Boulevard West,
Bronx, New York 10468-1589, USA }

\date{\today}
\begin{abstract}
Time dependence of the size and chirality of a ferromagnetic skyrmion in a Heisenberg model with the magnetic field on a square lattice has been studied analytically and numerically. The lattice and the magnetic field generate strong time dependence of the skyrmion chirality. Due to nonlinearity, the lattice alone also generates strong intrinsic damping that leads to the skyrmion collapse via the emission of spin waves. In the absence of the magnetic field the collapse is slow for a large skyrmion but it becomes exponentially fast in the presence of the Landau-Lifshitz damping when the field is turned on. Magnons emitted by a collapsing skyrmion must have a discrete spectrum due to the quantization of the skyrmion magnetic moment. 
\end{abstract}

\maketitle

\section{Introduction} \label{Intro}

Skyrmions in ferromagnetic films have been at the forefront of research in magnetism due to their interesting topological properties and potential for developing topologically protected data storage and information processing \cite{Leonov-NJP2016,Fert-Nature2017,Bogdanov2020,Luo2021}. Most of this research focuses on characteristics of stable skyrmions and their manipulation by spin-polarized currents or by other means. Less attention has been paid to the dynamics of unstable skyrmions. Experimental studies of these effects require different techniques \cite{Muckel2021} compared to those employed to visualize skyrmions \cite{Romming,McGray} because the corresponding timescales can be very short. However, they should be understood in processes of ultrafast data processing with skyrmions, and, therefore, deserve a separate investigation.

Skyrmions proliferated to condensed matter physics \cite{BP,Manton-book,Lectures} from nuclear physics \cite{SkyrmePRC58,Polyakov-book} due to the correspondence between the continuous spin-field model of the ferromagnetic exchange interaction and the $\sigma$-model of nuclear matter. Similar topological arguments that lead to skyrmions arise in Bose-Einstein condensates \cite{AlkStoNat01}, quantum Hall effect \cite{SonKarKivPRB93,StonePRB93}, anomalous Hall effect
\cite{YeKimPRL99}, liquid crystals \cite{WriMerRMR89} and graphene \cite{graphene}. 

In a pure two-dimensional (2D) field-theory model the energy of the skyrmion is independent of its size due to the scale invariance of the model. Unlike the relativistic field theory, however, the solid-state physics deals with atomic lattice that breaks the scale invariance of the model, making skyrmions unstable in the absence of interactions other than exchange. Early on, it was realized \cite{Bogdanov1989,Bogdanov94,Bogdanov-Nature2006} that skyrmions can be stabilized in materials with broken inversion symmetry by a combined effect of the magnetic field and Dzyaloshinskii-Moriya (DMI) interaction. However, many 2D ferromagnets and antiferromagnets (like, e.g, weakly interacting magnetic layers of parental compounds of high-temperature superconductors \cite{Klemm}) do not possess DMI and the question comes up how would topological defects like skyrmions, created in these materials by temperature or other means, evolve with time.

This question was initially addressed in a pure exchange model by Abanov and Pokrovsky \cite{Abanov} who accounted for the topological effect of the atomic lattice by ``punching a hole'' at the center of the skyrmion. They found that Landau-Lifshitz equation with dissipation provided collapse of the ferromagnetic skyrmion at the rate proportional to the damping constant and inversely proportional to the third power of the skyrmion size. This problem was later revisited for both ferro- and antiferromagnetic skyrmion by Cai et al. \cite{CCG-PRB2012} who derived lattice contribution to the spin Hamiltonian from a microscopic Heisenberg exchange interaction. They found that antiferromagnetic skyrmions collapsed much faster than ferromagnetic skyrmions. This was attributed to the suppression of the skyrmion collapse by the conservation of the angular momentum and the absence of inertia for the isotropic Heisenberg ferromagnet. The non-dissipative collapse of the ferromagnetic skyrmion on the lattice was attributed to the intrinsic damping through emission of spin excitations due to nonlinear dynamics. Quantum effects in the collapse of antiferromagnetic skyrmion in a pure exchange model have been studied in Ref.\ \onlinecite{AFM}. Theory of quantum collapse of a ferromagnetic skyrmion in the model with DMI and magnetic field was developed in Ref.\ \onlinecite{Amel2018}.

In this article we address the problem of the collapse of a ferromagnetic skyrmion in a centrosymmetric film in a more rigorous way. The collapse rate is obtained as a function of the magnetic field and damping analytically and numerically by solving the dynamics of one million spins on a square lattice. We show that the skyrmion collapse is determined by a subtle interplay of the effects of the magnetic field and the lattice. While the field alone does not lead to the skyrmion collapse in the absence of the Landau-Lifshits damping, the lattice, through nonlinearity, generates strong intrinsic damping that dominates the collapse of a small skyrmion.  

The article is organized as follows. Skyrmion dynamics in a continuous spin-field model is studied in Section \ref{Theory}. It begins with the effect of the lattice on the BP skyrmion in a pure 2D Heisenberg model, and then adds the magnetic field and dissipation to the model.  The section ends with considering quantum effects in skyrmion dynamics. Section \ref{Numerical} reports numerical studies of these effects in a discrete 2D spin model on a lattice and provides the comparison with analytical studies. Our results and their relevance to experiments are discussed in Section \ref{Discussion}. 

\section{Skyrmion dynamics in continuous spin-field models}\label{Theory}

\subsection{Ferromagnetic skyrmion in the exchange model on a lattice} \label{BP}
A continuous two-dimensional (2D) model with the Hamiltonian $\mathcal{H}  =  \frac{1}{2}JS^2\int dxdy\, \partial_{i}\mathbf{s} \cdot \partial_{i}\mathbf{s}$, describing Heisenberg nearest-neighbor exchange interaction of strength $J$ between three-component spins of length $S$,  has a skyrmion solution \cite{SkyrmePRC58} that in polar coordinates $(\phi, r)$ is given by \cite{BP,Polyakov-book,Lectures,Manton-book}
\begin{equation}
{\bf s}_{BP}=\left(\frac{2{\lambda} {r}\cos(\phi + \gamma)}{{r}^2+{\lambda}^{2}},\,\frac{2{\lambda} {r}\sin(\phi + \gamma)}{{r}^2+{\lambda}^{2}}\,, \frac{{\lambda}^2-{r}^{2}}{{\lambda}^2+{r}^{2}}\right).
\label{BP}
\end{equation}
It is commonly called the Belavin-Polyakov (BP) skyrmion. Here ${\bf s}$ is a unit vector, $\gamma$ is an arbitrary chirality angle, and $\lambda$ is an arbitrary scaling parameter associated with the size of the skyrmion. The uniform ferromagnetic order with ${\bf s} = -\hat{z}$ is assumed at infinity. Stability of the skyrmion is provided by the conservation of the topological charge $Q = \frac{1}{4\pi}\int dxdy\,{\bf s}\cdot(\partial_{x}{\bf s}\times\partial_{y}{\bf s})$. Here we consider the skyrmion of charge $Q = 1$. The energy of the BP skyrmion, $E_{BP} = 4\pi J$,  is independent of $\gamma$ and $\lambda$. 

When the 2D exchange model is studied on a square lattice of spacing $a$, the Hamiltonian, in the lowest order on $a$, acquires \cite{CCG-PRB2012} an additional term proportional to $a^2$,
\begin{equation}
\mathcal{H} =  JS^2\int
dxdy\, \left(\frac{1}{2}\partial_{i}\mathbf{s} \cdot \partial_{i}\mathbf{s} - \frac{a^2}{24}\partial^2_{i}\mathbf{s} \cdot \partial^2_{i}\mathbf{s}\right), 
\label{H-a}
\end{equation} 
At $\lambda \gg a$ Eq.\ (\ref{BP}) still provides a good approximation for the extremal solution but the second term in Eq.\ (\ref{H-a}) contributes $- {2\pi JS^2a^2}/({3\lambda^2})$ to the energy of the skyrmion. Consequently, on a lattice, smaller skyrmions have lower energy than larger skyrmions, which should not be surprising since the lattice violates the scale invariance of the continuous 2D Heisenberg model. 

Nevertheless, in the first approximation, the ferromagnetic skyrmion in a pure exchange model on a lattice would not collapse, because it has a nonzero total spin, 
\begin{equation}
{\Sigma}_{BP} = S\int dx dy \,(1 + \hat{z} \cdot {\bf s}_{BP})  = 4 \pi S \left(\frac{\lambda}{a}\right)^2\ln\left(\frac{L}{a}\right),
\label{Stot}
\end{equation}
that is conserved due to the invariance of the Hamiltonian (\ref{H-a}) with respect to rotations about the z-axis. (Here $L$ is a large-distance cutoff determined by the lateral dimension of the 2D lattice.) This situation is fundamentally different from an antiferromagnetic skyrmion that does not possess a nonzero total spin and collapses fast according to the equations of motion in excellent agreement with numerical simulations on a lattice \cite{CCG-PRB2012}. For a large ferromagnetic skyrmion, however, the situation is somewhat similar to that of a domain wall in a Heisenberg model with uniaxial magnetic anisotropy. If the field applied along the anisotropy axis, the wall is stationary due to the conservation of the angular momentum despite the fact that its motion would lower the energy. In Section \ref{Numerical} we will see that this argument breaks for a skyrmion on a lattice due to the additional nonlinearity of the problem brought by the lattice.

Introducing spherical coordinates $\Phi(x,y), \Theta(x,y)$ of the spin vector ${\bf s}$, the equations of motion for the size and chirality of the ferromagnetic skyrmion in this approximation can be obtained from the Lagrangian \cite{Lectures}
\begin{equation}
{\cal{L}} = \hbar S\int r dr d\phi  \,\dot{\Phi}(\cos\Theta + 1) - \mathcal{H}.
\label{Lag}
\end{equation}
For the BP skyrmion $\tan\Phi = s_y/s_x = \tan(\phi + \gamma)$, yielding $\Phi = \phi + \gamma$,  $\dot{\Phi} = \dot{\gamma}$. Assuming that $\gamma$ is independent of coordinates, one obtains from Eq.\ (\ref{Lag})
\begin{equation}
{\cal{L}} = 4\pi \hbar S \dot{\gamma}\left(\frac{\lambda}{a}\right)^2 \ln\left(\frac{L}{\lambda}\right) +  \frac{2\pi }{3}JS^2\left(\frac{a}{\lambda}\right)^2.
\end{equation}
The Euler-Lagrange equations are
\begin{equation}
\frac{d}{dt} \frac{\partial {\cal{L}}}{\partial \dot{\gamma}} = \frac{\partial {\cal{L}}}{\partial {\gamma}}, \qquad \frac{d}{dt} \frac{\partial {\cal{L}}}{\partial \dot{\lambda}}=\frac{\partial {\cal{L}}}{\partial {\lambda}}. 
\label{EL}
\end{equation}
From the first of these equations 
\begin{equation}
\frac{d}{dt} \left(\lambda^2\ln\frac{L}{\lambda}\right) = 0, \quad \lambda = {\rm const}
\end{equation}
With the account of this condition, the second equation in (\ref{EL}) gives
\begin{equation}
\hbar\dot{\gamma}= \frac{JS}{6\ln(L/\lambda\sqrt{e})}\left(\frac{a}{\lambda}\right)^4, \quad \gamma =  \frac{JSt}{6\hbar\ln(L/\lambda\sqrt{e})}\left(\frac{a}{\lambda}\right)^4.
\label{gamma-t}
\end{equation}

Thus, the presence of the lattice makes the chirality $\gamma$ of the skyrmion in Eq.\ (\ref{BP}) change linearly with time at the rate that rapidly goes down as $1/\lambda^4$ with increasing the skyrmion size.  The size $\lambda$ of the skyrmion remains constant in this approximation in accordance with the conservation of  the total spin of the skyrmion given by Eq.\ (\ref{Stot}). This should not be surprising because the greater the skyrmion size the better is the scale-invariant continuous approximation in which the skyrmion of any size is stable. 

\subsection{Effect of the magnetic field}\label{field}

Let us now include into the problem the external magnetic field $H$ opposite to the z-axis. In practice, its presence is needed to provide the condition ${\bf s} = -\hat{z}$ at infinity. Zeeman contribution, $-g\mu_B \Sigma_{BP} H$, of this negative field to the Lagrangian (with the parameter $H$ being positive in all equations) changes it to
\begin{equation}
{\cal{L}} = 4\pi \hbar S \left(\dot{\gamma}-\frac{g\mu_B H}{\hbar}\right)\left(\frac{\lambda}{a}\right)^2 \ln\left(\frac{L}{\lambda}\right) +  \frac{2\pi }{3}JS^2\left(\frac{a}{\lambda}\right)^2 .
\end{equation}
The solution for $\gamma$ changes accordingly:
\begin{eqnarray}
\hbar\dot{\gamma} & = & g\mu_B H + \frac{JS}{6\ln(L/\lambda\sqrt{e})}\left(\frac{a}{\lambda}\right)^4  \label{gamma-dot} \\
\gamma & =  & \frac{g\mu_B H}{\hbar} t + \frac{JS}{6\hbar\ln(L/\lambda\sqrt{e})}\left(\frac{a}{\lambda}\right)^4 t.
\end{eqnarray}
This can also be seen from the fact that, mathematically, adding the term $-g\mu_B ({\bf s} \cdot {\bf H}) $ to the Hamiltonian density is equivalent to switching to the frame of reference rotating with the angular velocity ${\bm \Omega} = (g\mu_B {\bf H}/\hbar)$. The corresponding transformation of the components of the spin field adds $\Omega t$ to the chirality $\gamma$ in Eq.\ (\ref{BP}). Less obvious is the fact that the effect of the crystal lattice is equivalent to the effect of the fictitious magnetic field determined by the exchange and the size of the skyrmion.

\subsection{Skyrmion collapse in the presence of dissipation} \label{dissipation}

In practice, the spin dynamics in solids is always damped. The effect of damping can be described by the dissipation function \cite{Brown}
\begin{equation}
F = \frac{1}{2}\hbar \eta_0 S^2 \int rdrd\phi \,\left(\frac{\partial {\bf s}}{\partial t}\right)^2,
\label{F}
\end{equation}
where $\eta_0 \ll 1$ is the damping parameter that coincides with the damping factor in the Landau-Lifshitz equation \cite{Lectures}. We compute $F$ using the BP solution (\ref{BP}). Since without damping $\lambda \approx {\rm const}$, its time derivative due to damping must be proportional to $\eta_0$ in this approximation. On the contrary, the time derivative of $\gamma$ given by Eq.\ (\ref{gamma-dot}) is independent of $\eta_0$. Consequently, it must dominate the dissipation function. Computing its contribution to the time-derivative of the spin field, 
\begin{equation}
\left(\frac{\partial {\bf s}}{\partial t}\right)^2 = \left(\frac{\partial {s}_{BP,x}}{\partial t}\right)^2 + \left(\frac{\partial {s}_{BP,y}}{\partial t}\right)^2 = \frac{4\lambda^2 r^2}{(r^2 + \lambda^2)^2} \left(\frac{d\gamma}{dt}\right)^2,
\end{equation}
and integrating in Eq.\ (\ref{F}) we obtain
\begin{equation}
F = 4\pi \hbar \eta_0 S^2 \dot{\gamma}^2 \left(\frac{\lambda}{a}\right)^2 \ln\left(\frac{L}{\lambda\sqrt{e}}\right).
\end{equation}

The first Euler-Lagrange equation in Eq.\ (\ref{EL}) must now be replaced with
\begin{equation}
\frac{d}{dt} \frac{\partial {\cal{L}}}{\partial \dot{\gamma}} = \frac{\partial {\cal{L}}}{\partial {\gamma}} - \frac{\partial F}{\partial \dot{\gamma}}\,.
\end{equation}
This gives
\begin{equation}
\frac{d\lambda}{dt} = -\eta_0 S\lambda \dot{\gamma}. 
\label{lambda-dot-gen}
\end{equation}
Substituting here $\dot{\gamma}$ of Eq.\ (\ref{gamma-dot})  we obtain
\begin{equation}
\frac{d\lambda}{dt} = -\frac{\eta_0 S}{\hbar} \lambda \left[ g\mu_B H  + \frac{JS}{6\ln(L/\lambda\sqrt{e})} \left(\frac{a}{\lambda}\right)^4\right].
\label{lambda-dot}
\end{equation}

According to Eq.\ (\ref{lambda-dot}) the collapse of the skyrmion is markedly different for $\lambda \gg \lambda_1(H)$ and for $\lambda \ll \lambda_1(H)$, where $\lambda_1$ is solution of the equation
\begin{equation}
H = \frac{JS}{6g\mu_B \ln(L/\lambda_1 \sqrt{e})}\left(\frac{a}{\lambda_1}\right)^4.
\label{lambda-1}
\end{equation}
At $\lambda \gg \lambda_1$ it is dominated by the magnetic field and is exponential in time, 
\begin{equation}
\lambda = \lambda_0 \exp\left(-\frac{\eta_0 g\mu_B S H}{\hbar} t\right),
\label{exp}
\end{equation}
with the time constant inversely proportional to the damping constant and the field, $t_H = \hbar/(\eta_0 g\mu_B S H)$.  In the opposite limit of  $\lambda \ll \lambda_1$, the solution of Eq.\ (\ref{lambda-dot})  is
\begin{equation}
\left(\frac{\lambda}{a}\right)^4 \ln\left(\frac{L}{\lambda e^{1/4}}\right)  = \left(\frac{\lambda_1}{a}\right)^4 \ln\left(\frac{L}{\lambda_1 e^{1/4}}\right) - \frac{2\eta_0 J S^2}{3\hbar} t,
\label{small-lambda}
\end{equation}
and the rest of the collapse takes time
\begin{equation}
t_c = \frac{3\hbar}{2\eta_0 J S^2}\left(\frac{\lambda_1}{a}\right)^4 \ln\left(\frac{L}{\lambda_1 e^{1/4}}\right)
\label{tc}
\end{equation}
to complete. This last formula agrees with the result of Ref.\ \onlinecite{Abanov} for the dependence of the collapse time (obtained by a different method) on the initial skyrmion size, $\lambda_1 = \lambda_0$, in the absence of the magnetic field. 

It is easy to see that $t_c =t_H/4$. Consequently, as long as the field in Eq.\ (\ref{lambda-1}) is sufficiently large to provide $\lambda_1 < \lambda_0$, the lifetime  of the skyrmion is determined by $t_H$ which is independent of the initial skyrmion size and is inversely proportional to $H$. For weaker fields the lifetime of the skyrmion is given by $t_c$ of Eq.\ (\ref{tc}) with $\lambda_1$ replaced by $\lambda_0$.

\subsection{Quantum effects}\label{quantum}

In quantum mechanics the total spin of the skyrmion (\ref{Stot}) is quantized, which leads to the quantization of $\lambda$. The distance in $\lambda$ between two adjacent energy levels, 
\begin{equation}
\Delta \lambda = \frac{a^2}{8\pi S \lambda  \ln({L}/{\lambda \sqrt{e}})},
\label{Deltalambda}
\end{equation}
follows from the condition $\Delta \Sigma_{BP} = 1$.  With the sum of Zeeman and lattice energies given by
\begin{equation}
E = 4\pi g\mu_B S H \left(\frac{\lambda}{a}\right)^2 \ln\left(\frac{L}{\lambda}\right) -  \frac{2\pi }{3}JS^2\left(\frac{a}{\lambda}\right)^2 
\end{equation}
the energy difference between the adjacent levels is
\begin{equation}
\Delta E \equiv \hbar \omega =  \frac{dE}{d\lambda}\Delta \lambda =  {g\mu_B H} + \frac{JS}{6\ln(L/\lambda\sqrt{e})}\left(\frac{a}{\lambda}\right)^4, 
\label{omega}
\end{equation}
where we have used Eq.\ (\ref{Deltalambda}) for $\Delta \lambda$. Remarkably, $\omega$ is exactly the frequency of the precession of skyrmion spins given by $\dot{\gamma}$ of Eq.\ (\ref{gamma-dot}).  It suggests that it is the frequency of the spin waves emitted by the collapsing skyrmion when its total spin decreases from $\Sigma_{BP}$ to zero in jumps of $\Delta \Sigma_{BP} = 1$. Since each magnon carries spin 1, the emission of magnons occurs at a rate 
\begin{equation}
\Gamma = - \frac{d \Sigma_{BP}}{dt} = -\frac{8\pi S \lambda}{a^2} \left(\frac{d \lambda}{dt}\right) \ln\left(\frac{L}{\lambda \sqrt{e}}\right).
\end{equation}
Substituting here $d\lambda/dt$ of Eq.\ (\ref{lambda-dot}), one obtains
\begin{eqnarray}
\Gamma & = & 8\pi \eta_0 S^2 \left[ \frac{g\mu_B H}{\hbar}\left(\frac{\lambda}{a}\right)^2  + \frac{JS}{6\hbar\ln(L/\lambda\sqrt{e})} \left(\frac{a}{\lambda}\right)^2\right] \nonumber \\
& \times & \ln\left(\frac{L}{\lambda \sqrt{e}}\right).
\end{eqnarray}
Notice that $\lambda$ in the above formulas is quantized, so that magnons emitted by the collapsing skyrmion must have a discrete spectrum determined by the quantization of $\Sigma_{BP}$ in Eq.\ (\ref{Stot}).

\section{Numerical study of skyrmion dynamics on a lattice}\label{Numerical}

In this section we will study dynamical effects considered in the previous sections by introducing microscopic Hamiltonian on a square lattice,
\begin{equation} 
\mathcal{H} =  -\frac{S^{2}}{2}\sum_{ij}J_{ij}\textbf{s}_{i}\cdot\textbf{s}_{j}  - g\mu_B S\textbf{H}\cdot\sum_{i}\textbf{s}_{i} ,
\label{energy-discrete}
\end{equation}
and solving numerically the Landau-Lifshitz equations for the spins:
\begin{equation}
\hbar\dot{\textbf{s}}_{i}=[\textbf{s}_{i}\times(g\mu_B\textbf{H}_{\mathrm{eff},i}]-\eta_0[\textbf{s}_{i}\times(\textbf{s}_{i}\times g\mu_B\textbf{H}_{\mathrm{eff},i})],
\end{equation}
where  $g\mu_{0}S\textbf{H}_{\mathrm{eff},i}=-\partial\mathcal{H}/\partial\textbf{s}_{i}$.

\subsection{Numerical method}\label{method}

Numerical results are obtained by computing the time evolution of an array of spins undergoing LL dynamics in a 2D lattice. A BP skyrmion of size $\lambda_0$ is initialized on a $3 \times N_x \times N_y$ array according to Eq.\ (\ref{BP}).  Its time evolution is obtained using a fourth-order Runge-Kutta ODE solver, with the timestep $0.2$ in units of $\hbar/J$. We used free boundary conditions and checked that the total energy of the spin system was conserved with high accuracy. The computation runs until the skyrmion collapses, which occurs when the topological charge drops from $Q = 1$ to $Q = 0$. Throughout the computation, the topological charge, total energy, chirality, and the effective size of the skyrmion are measured. The latter is given by \cite{Amel2018}
\begin{equation}
\frac{\lambda_{\rm eff}^2}{a^2} = \frac{n-1}{2^n \pi}\sum_i (s_{z,i} +1)^n.
\end{equation}
For BP skyrmions, $\lambda_{\rm eff} = \lambda$ for any $n$. We use $n=4$. 

The chirality is measured by noting that a cylindrically symmetric spin texture can be written as $(\sin\Theta \cos\Phi, \sin\Theta \sin\Phi, \cos\Theta)$, with $\Phi = \phi + \gamma$, in terms of the spherical angles $\Theta$ and $\Phi$ of the spin vector, polar angle $\phi$ in the XY plane, and the chirality angle $\gamma$. Along the line $\phi = 0$ and far from the skyrmion center, $\gamma$ can be computed as
\begin{equation}
\gamma = \cos^{-1}\left(\frac{s_x}{\sqrt{1-s_z^2}}\right).
\end{equation}
In this way one can compute the chirality from a single chosen point in the array that is sufficiently far from the skyrmion center.  We use the point $x = 20$ along the line $\phi = 0$. 

The lattice size has been chosen to balance the computation time with the desirable accuracy. For the conservative model, computation was performed on a $1024 \times 1024$ lattice. For the damped model we used a $512 \times 512$ lattice. 

The numerical program is written in Julia and is available on GitHub. Computations are completed by parallelizing different collapse events on Lehman College's 40-node computing cluster.

\begin{figure}[h]
\hspace{-0.2cm}
\includegraphics[width=8.7cm]{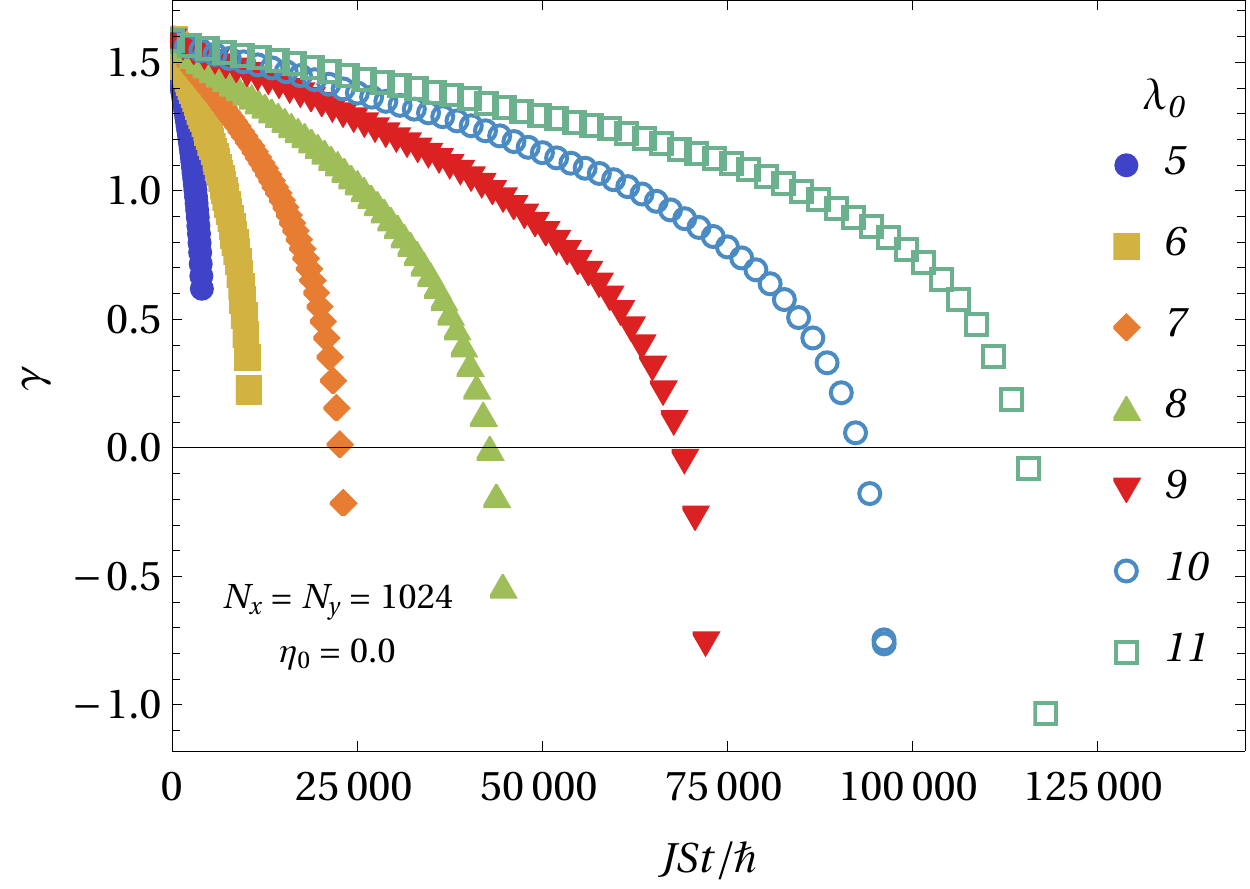} 
\caption{Time dependence of the chirality of a ferromagnetic skyrmion on the lattice.}
\label{gamma-time}
\end{figure}
\begin{figure}[h]
\hspace{-0.2cm}
\includegraphics[width=8.7cm]{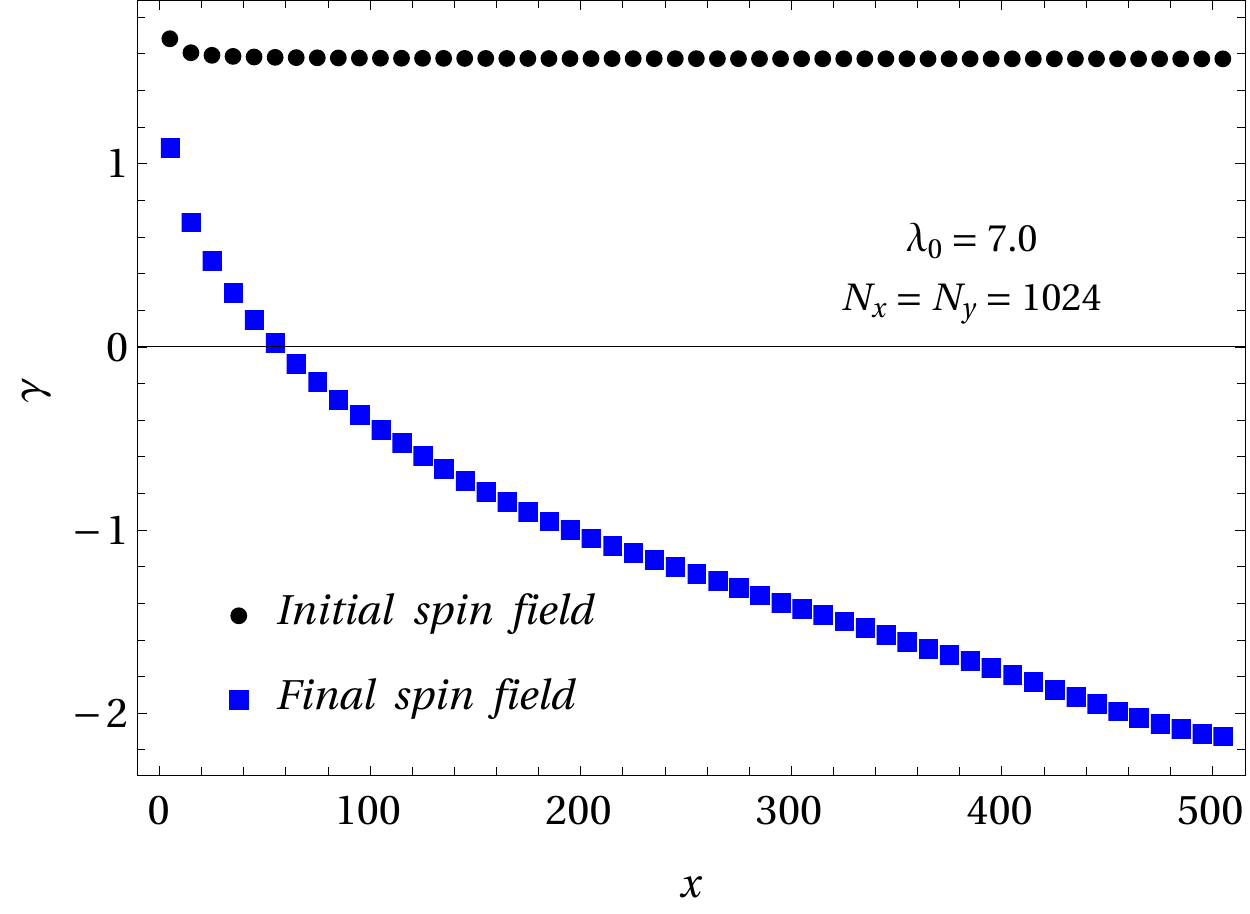} 
\caption{Chirality of the skyrmion as a function of the position in the lattice, $n_x$, for the initial and final spin fields. Values are computed along the line $\phi = 0$.}
\label{gamma-n}
\end{figure}
\begin{figure}[h]
\hspace{-0.2cm}
\includegraphics[width=8.7cm]{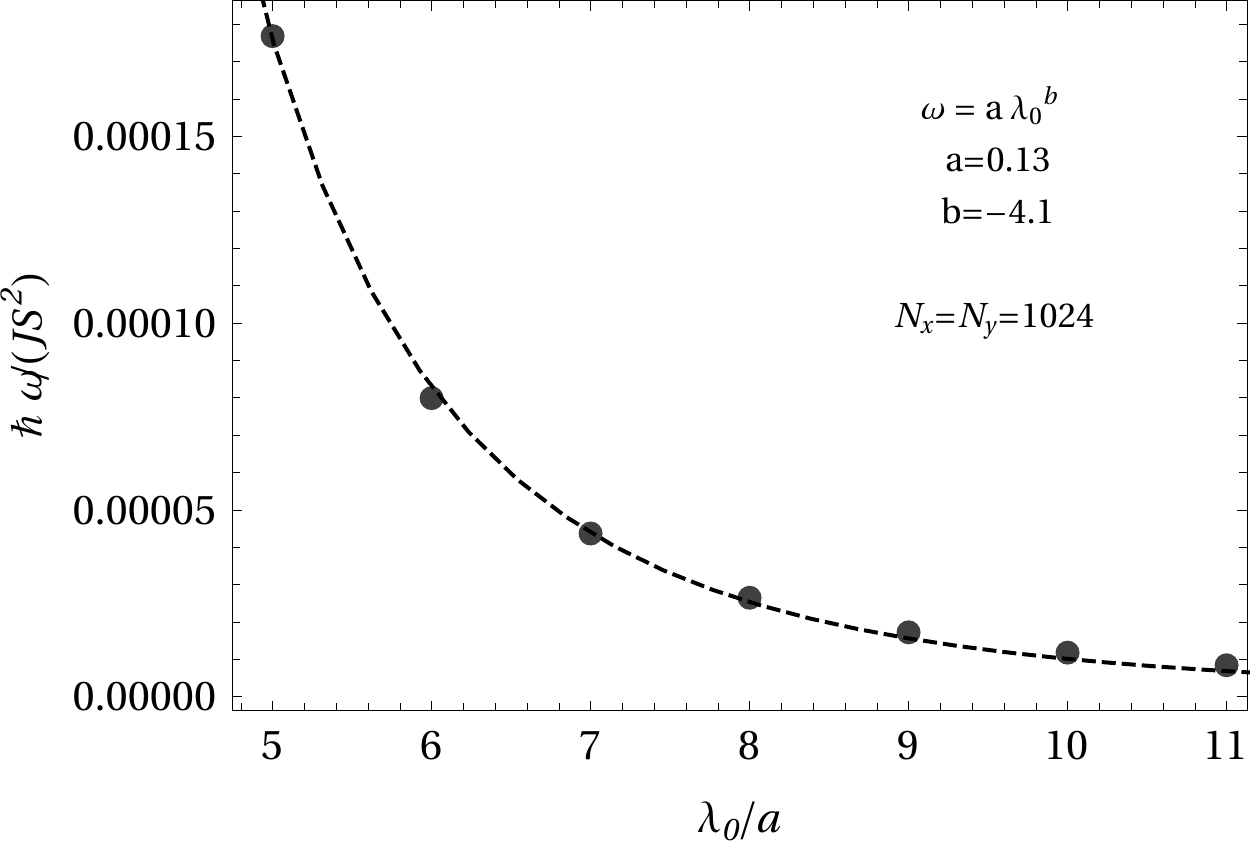} 
\caption{Time derivative of the chirality, $\omega = \dot{\gamma}$, versus the initial size of the skyrmion.}
\label{gamma-size}
\end{figure}

\subsection{Skyrmion collapse in the absence of the magnetic field}\label{num-BP}

\begin{figure}[h]
\hspace{-0.2cm}
\includegraphics[width=8.7cm]{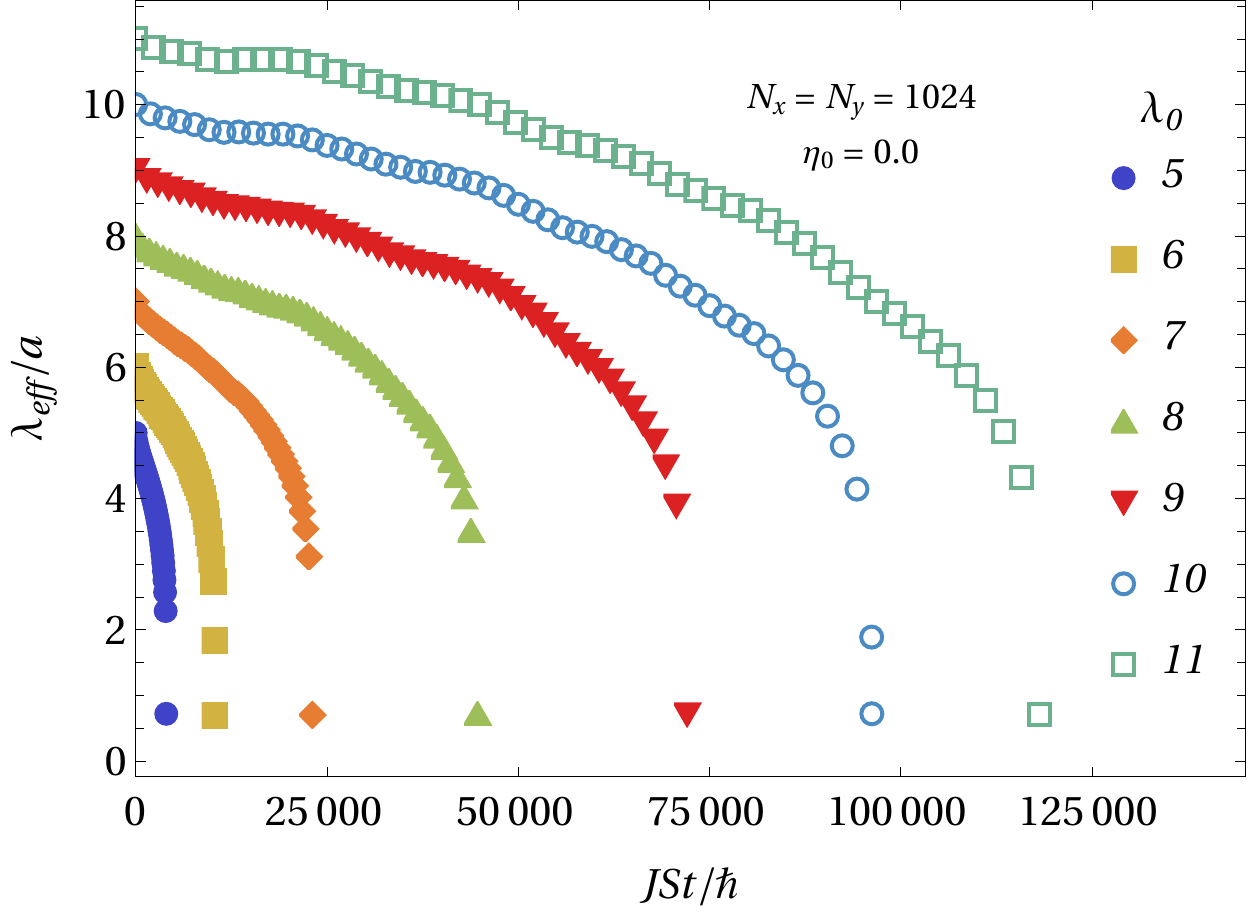} 
\caption{Numerical data for the collapse of the ferromagnetic skyrmion of different initial size in a pure exchange model with no magnetic field and no LL damping, on a $1024 \times 1024$ square lattice.}
\label{collapse}
\end{figure}
We first consider skyrmion collapse in a generic exchange model on a lattice with no magnetic field and no LL damping. Analytical approximation that uses the BP shape of the skyrmion does not provide its collapse in the absence of the LL damping, although for large skyrmions the lattice generates linear dependence of the skyrmion chirality on time given by Eq.\ (\ref{gamma-t}).  This prediction is confirmed by our numerical studies, see. The chirality initially changes linearly with time. The rate of change accelerates as the skyrmion becomes close to the collapse. The failure of the BP approximation (valid for $\lambda \gg a$) at the final stage of the collapse can additionally be observed by computing the chirality as a function of the position in the XY plane as time progresses, see Fig.\ \ref{gamma-n}. Eq.\ (\ref{gamma-t}) predicts that the coefficient of proportionality in the linear time dependence of $\gamma$ is, up to a logarithm, inversely proportional to the fourth power of the skyrmion size. This prediction is also confirmed by the numerical experiment, see Fig.\ \ref{gamma-size}. 

Numerical results for the time dependence of the skyrmion size on a $1024 \times 1024$ square lattice are shown for a few initial sizes in Fig.\ \ref{collapse}. We see that in the generic Heisenberg model on a lattice the ferromagnetic skyrmion does collapse at $\eta_0 = 0$, though at a rate that is orders of magnitude slower than the collapse of the antiferromagnetic skyrmion \cite{CCG-PRB2012,AFM}. This must be influenced by 1) conservation of the total spin, which is not a factor in an antiferomagnet where it is zero, and 2) the absence of the inertia (skyrmion mass) \cite{Psaroudaki-PRX2017,Kravchuk-PRB2018} that drives the skyrmion dynamics in an antiferromagnet but is absent in a ferromagnet. 

\begin{figure}[h]
\hspace{-0.2cm}
\includegraphics[width=8.7cm]{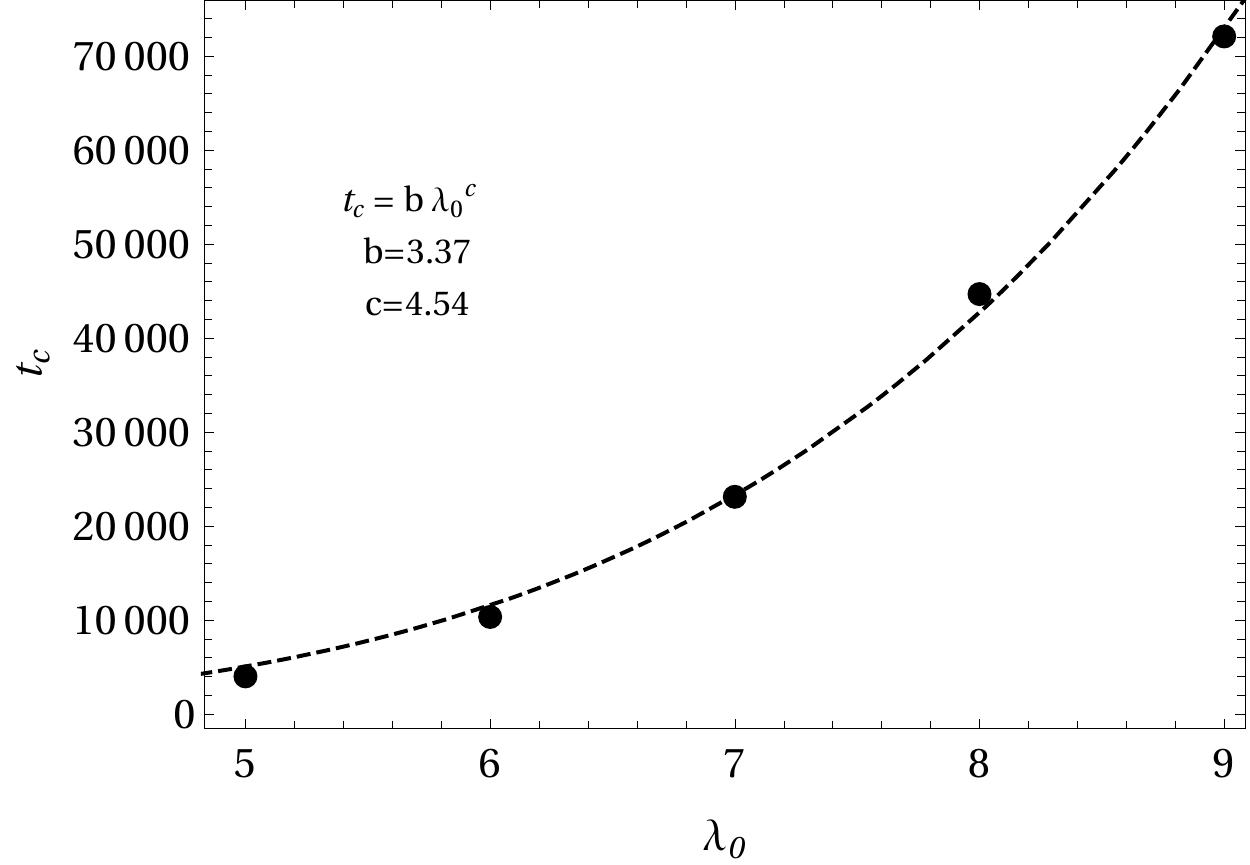} 
\caption{Skyrmion collapse time vs initial size at zero field.}
\label{tc-size}
\end{figure}
\begin{figure}[h]
\hspace{-0.2cm}
\includegraphics[width=8.7cm]{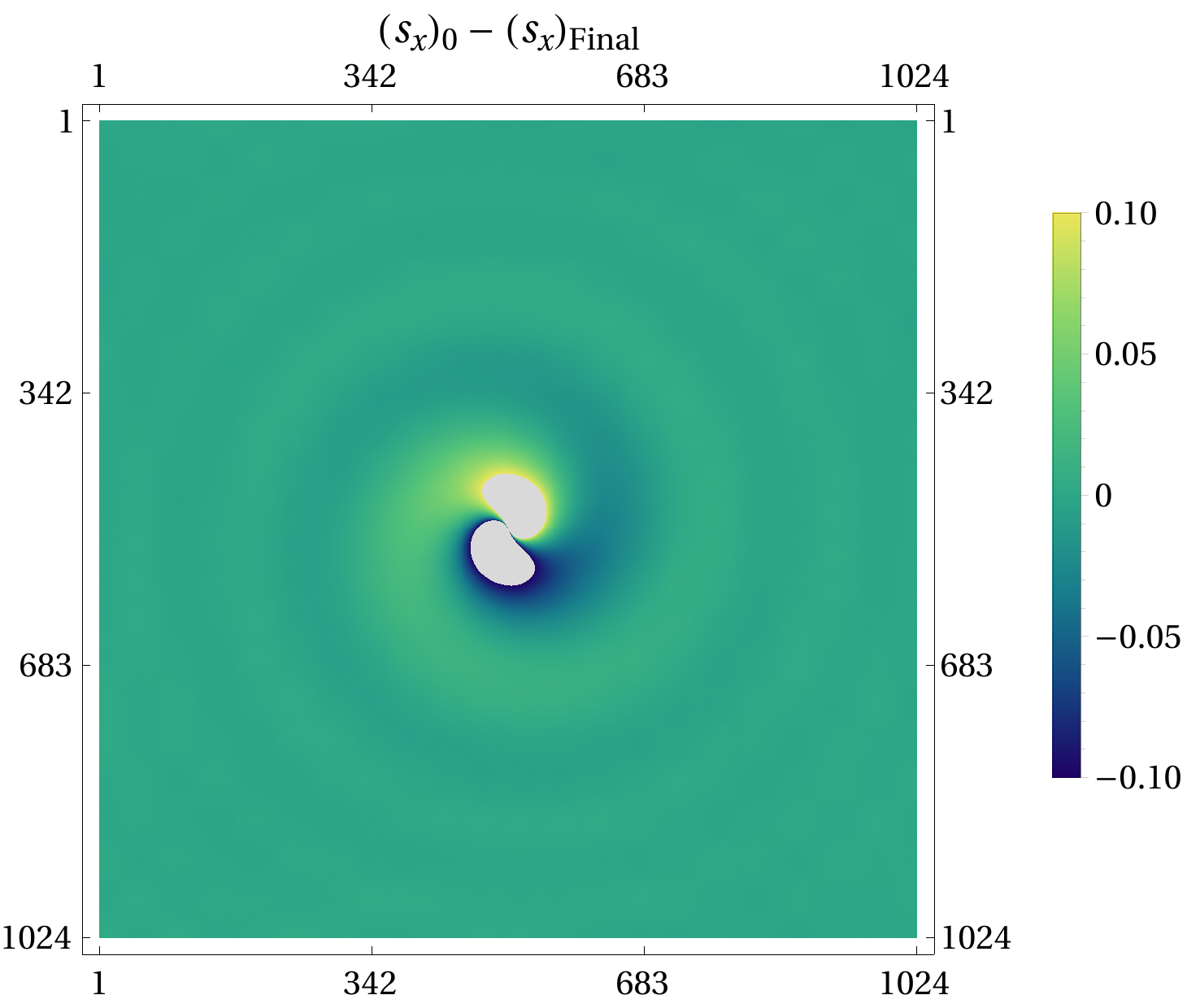} 
\caption{A snapshot of the collapse of the ferromagnetic skyrmion on a lattice. Initial spin density subtracted from the spin density of a skyrmion about to collapse. The plot range is limited to more clearly visualize spin waves. Gray areas demarcate values above and below the plot range in the legend on the right.}
\label{SW}
\end{figure}
Time-dependence of the skyrmion size during the collapse is shown in Fig.\ \ref{collapse}. The collapse time as a function of the initial size roughly follows the prediction of Eq.\ (\ref{tc}), see Fig.\ \ref{tc-size}. Remarkably, we find that in the absence of the LL damping the collapse rate is close to the law given by Eq.\  (\ref{lambda-dot}) at $H = 0$, with $\eta_0$ replaced by the intrinsic damping  $\eta_{in}$.  This is confirmed by fitting the collapse time as function of the total damping (LL + intrinsic), see Eq.\ (\ref{tc}), with $t_c \propto (\eta_{in} + \eta_0)^{-1}$ at a fixed $\lambda_0$. For small skyrmions we find $\eta_{in} \sim 0.4$. The observed large intrinsic dissipation must be due to the additional nonlinearity introduced by the lattice on top of the nonlinearity of the continuous BP model. It apparently generates breathing of the collapsing skyrmion accompanied by the emission of spin waves, see Fig.\ (\ref{SW}). 

The number of full rotations of the in-plane spin components during the contraction of the skyrmion from the initial size $\lambda_0$ is of order $(t_H + t_c)\dot{\gamma}$. According to the theory, it is inversely proportional to the damping factor that contains LL and intrinsic contributions. At small LL damping it is dominated by the intrinsic damping, making the number of rotations of order unity for the observed value of $\eta_{in} \sim 0.4$. This agrees with the numerical data presented in  Fig.\  \ref{gamma-time}. It also explains why it is difficult to develop full analytical theory of the collapse of the ferromagnetic skyrmion on a lattice: Large values of the effective damping factor $\eta_{in}$ due to the lattice makes its effect on the BP skyrmion nonperturbative. 

\subsection{Skyrmion collapse in the presence of the field}\label{num-field}

\begin{figure}[h]
\hspace{-0.2cm}
\includegraphics[width=8.7cm]{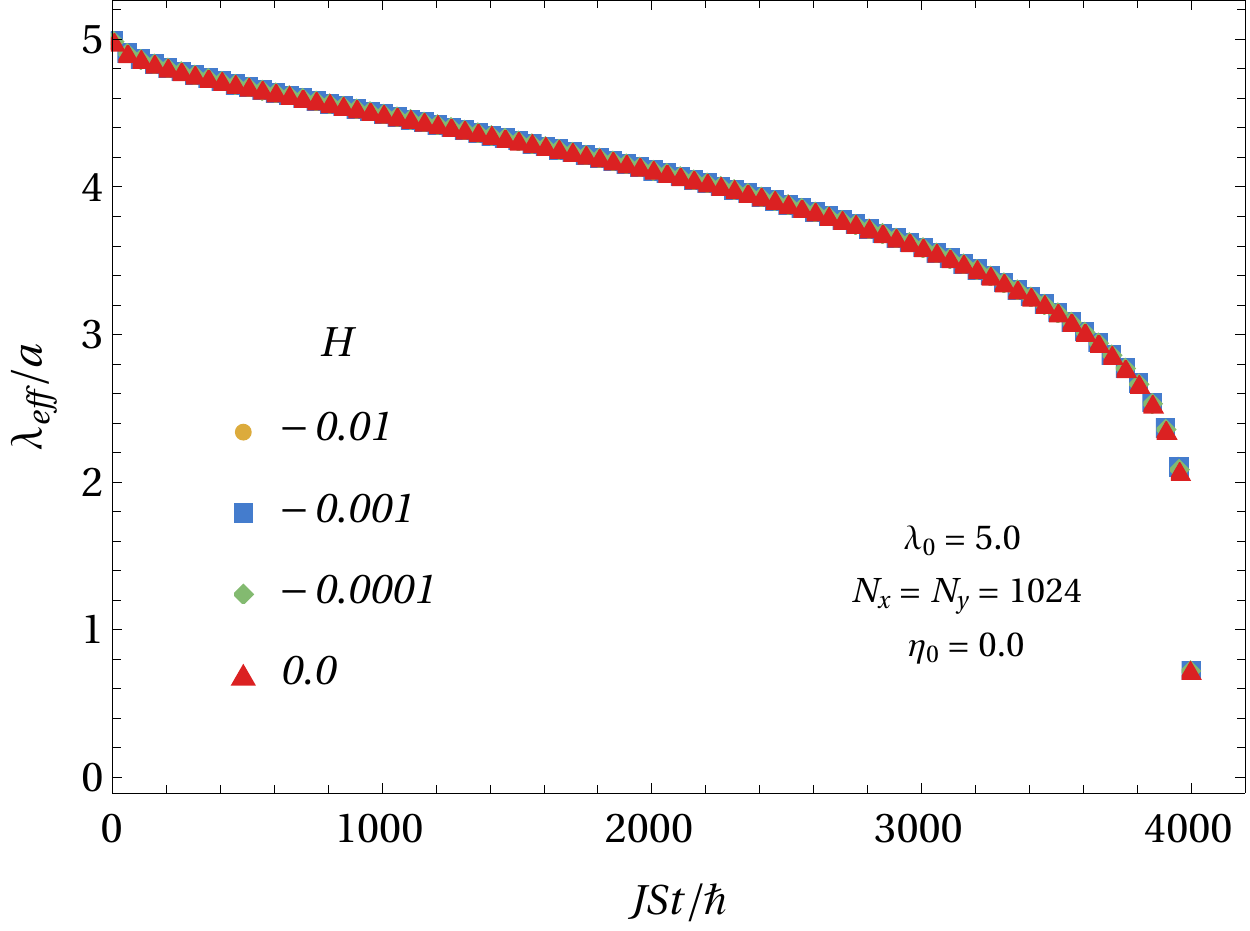} 
\caption{Collapse of the ferromagnetic skyrmion on a $1024 \times 1024$ square lattice in a pure exchange model with no damping in the presence of the magnetic field. Independence of $H$ is apparent.}
\label{l-t}
\end{figure}
We now include the magnetic field into the problem. Interestingly, it has no effect on the collapse of the ferromagnetic skyrmion at $\eta_0 = 0$, see Fig.\ \ref{l-t}. In the absence of the LL damping it is entirely determined by the intrinsic dissipation due to the lattice. This suggests the following modification of Eq.\ (\ref{lambda-dot}):
\begin{equation}
\frac{d\lambda}{dt} = -\frac{ S}{\hbar} \lambda \left[ \eta_0g\mu_B H  + \frac{(\eta_0 + \eta_{in})JS}{6\ln(L/\lambda\sqrt{e})} \left(\frac{a}{\lambda}\right)^4\right].
\label{lambda-dot-in}
\end{equation}
It leads to the modification of the formula (\ref{lambda-1}) that determines skyrmion size $\lambda_1$ separating two regimes of the collapse,
\begin{equation}
H = \frac{JS}{6\mu_B \ln(L/\lambda_1\sqrt{e})}\left(1 + \frac{\eta_{in}}{\eta_0}\right)\left(\frac{a}{\lambda_1}\right)^4.
\end{equation}
At a fixed $H$, the value of $\lambda_1$ becomes roughly greater by a factor $(1 + \eta_{in}/\eta_0)^{1/4}$. Eq.\ (\ref{exp}) for the time dependence of the skyrmion size at $\lambda \gg \lambda_1$ does not change but in Eqs.\ (\ref{small-lambda}) and (\ref{tc}), describing the collapse at $\lambda \ll \lambda_1$, the LL damping constant $\eta_0$ has to be replaced with $\eta_0 + \eta_{in}$.

\begin{figure}[h]
\hspace{-0.2cm}
\includegraphics[width=8.7cm]{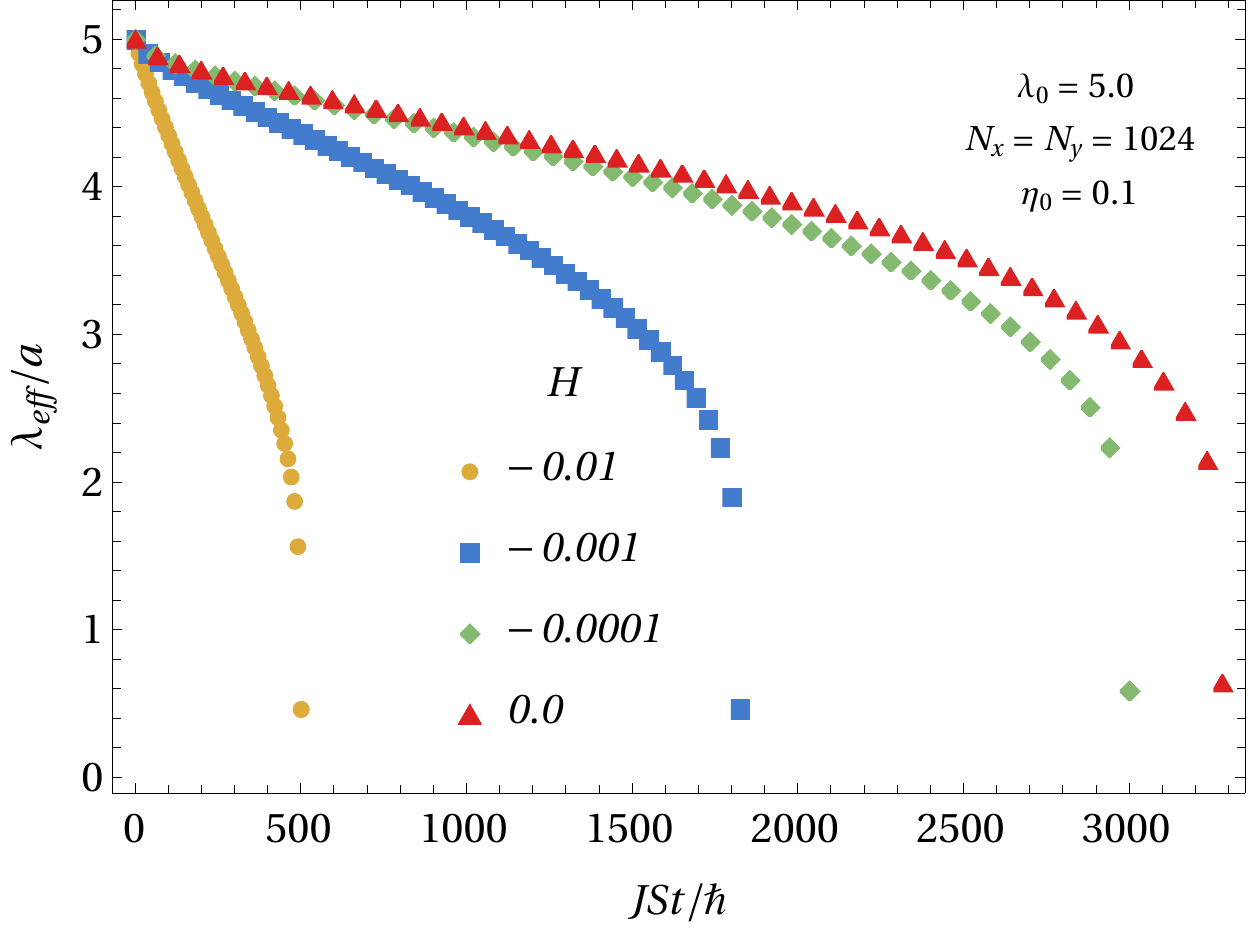} 
\caption{Collapse of the ferromagnetic skyrmion on a $1024 \times 1024$ square lattice in a pure exchange with the magnetic field and the LL damping $\eta_0 = 0.1$.}
\label{lamda-H-eta}
\end{figure}
\begin{figure}[h]
\hspace{-0.2cm}
\includegraphics[width=8.7cm]{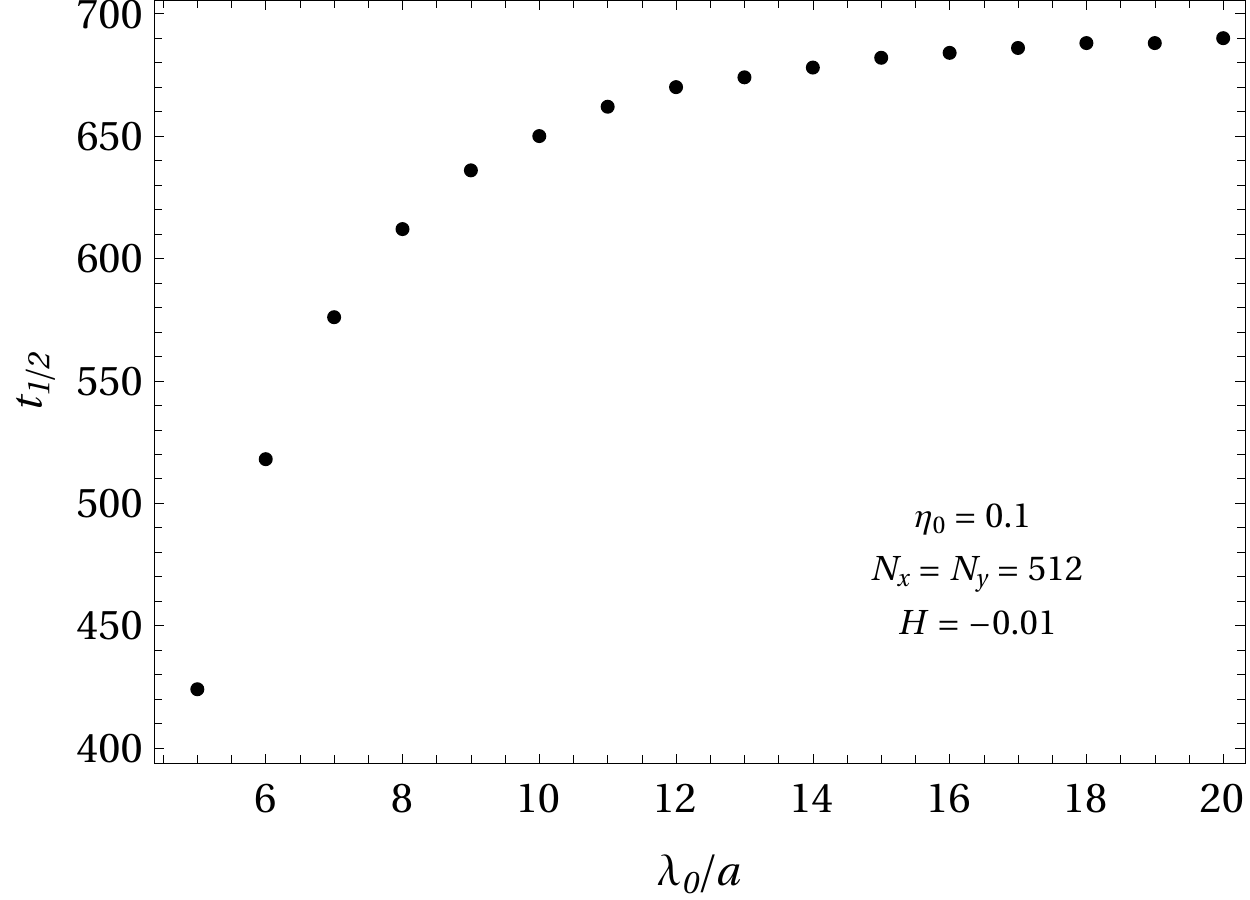} 
\caption{The dependence of half-life of the collapsing skyrmion on the intial size.}
\label{halflife}
\end{figure}
We test these predictions by turning on the LL damping, see Fig.\ \ref{lamda-H-eta}. In accordance with Eq.\ (\ref{lambda-dot-in}), the collapse begins to depend strongly on $H$ as soon as the two terms in that equation become comparable. At a fixed $H$ the switching between the two regimes discussed above occurs on decreasing $\lambda$. When the first term dominates, the collapse is exponential in time, with the skyrmion half-lifetime given by $t_{1/2}= t_H \ln(2)$, where $t_H = \hbar/(\eta_0 g\mu_B S H)$ as suggested by Eq.\ (\ref{exp}). Notice that according to Eq.\ (\ref{lambda-dot-in}) the collapse time of a sufficiently large skyrmion is dominated by the exponential regime, with the half-life independent of the initial size. This is confirmed by our numerical data presented in Fig.\ (\ref{halflife}) that show the tendency of $t_{1/2}$ to become independent of $\lambda_0$ on increasing the initial size of the skyrmion.

\section{Discussion}\label{Discussion}

We have studied analytically and numerically the collapse of the ferromagnetic skyrmion in a centrosymmetric atomic lattice. This problem is important for understanding skyrmion dynamics due to the effort to develop techniques of writing and deleting skyrmions in a magnetic film as a path towards topologically protected magnetic memory \cite{Zhang2020}. It is markedly different from the problem of the collapse of an antiferromagnetic skyrmion due to the absence of inertia in the ferromagnetic model. 

We find that in the absence of stabilizing interactions the lifetime of the ferromagnetic skyrmion depends on a subtle interplay between the effects of the crystal lattice and the magnetic field. In the analytical model, skyrmion collapse requires damping \cite{Abanov}. It is dominated by the dynamics of spins due to the fast change of the chirality angle with time, induced by the lattice and by the field. In the absence of the LL damping, the magnetic field has no effect on the skyrmion collapse. On the contrary, the lattice generates an effective intrinsic damping via nonlinearity that is equivalent to the effect of the LL dissipation with a damping factor of about 0.4. 

As the result, the rate of the contraction of the skyrmion size is determined by two contributions: the contribution of the lattice that is proportional to the sum of the LL damping and the intrinsic damping, and the contribution of the field that is proportional solely to the LL damping. These two contributions have different dependence on the skyrmion size. While the field contribution to the rate is proportional to the skyrmion size, the lattice contribution is, up to a logarithm, inversely proportional to the third power of the size. 

Consequently, in the presence of the field, sufficiently large skyrmions collapse exponentially with time, with the half-life inversely proportional to the field and independent of the initial size. The collapse time of a small skyrmion is proportional, up to a logarithm, to the fourth power of its size. The borderline between the two regimes depends on the field and the LL damping. In all cases the typical collapse time is below one nanosecond but can be easily resolved in experiment. 

Quantization of the magnetic moment of the skyrmion must lead to the quantization of the energy of magnons emitted by collapsing skyrmions. In experiments where skyrmions and antiskyrmions are nucleated by thermal fluctuations due to elevated temperature, or by other means, the quantization of the spectrum of magnons generated by the collapse of the skyrmions, may be detectable in the magnetic noise. Such spectroscopy of skyrmions would be similar to the optical spectroscopy of atoms of a hot substance.

\section{Acknowledgements}

This work has been supported by the Grant No. DE-FG02-93ER45487 funded by the U.S. Department of Energy, Office of Science.


\begin{thebibliography}{10}

\bibitem{Leonov-NJP2016} A. O. Leonov, T. L. Monchesky, N. Romming, A. Kubetzka, A. N. Bogdanov, and R. Wiesendanger, The properties of
isolated chiral skyrmions in thin magnetic films, New Journal of Physics \textbf{18}, 065003-(16) (2016).

\bibitem{Fert-Nature2017} A. Fert, N. Reyren, and V. Cros, Magnetic skyrmions: advances in physics and potential applications, Nature Reviews Materials \textbf{2}, 17031-(15) (2017).

\bibitem{Bogdanov2020}
A. N. Bogdanov and C. Panagopoulos, Physical foundations and basic properties of magnetic skyrmions, Nature Review Materials {\bf 2}, 492-498 (2020).

\bibitem{Luo2021}
S. Luo and L. You, Skyrmion devices for memory and logic applications, APL Materials {\bf 9}, 050901-(11) (2021). 

\bibitem{Muckel2021}
F. Muckel, S. von Malottki, C. Holl, B. Pestka, M. Pratzer, P. F. Bessarab, S. Heinze, and M. Morgensten, Experimental identification of two distinct skyrmion collapse mechanisms, Nature Physics {\bf 17}, 395-402 (2021). 

\bibitem{Romming}
N. Romming, C. Hanneken, M. Menzel, J. E. Bickel, B. Wolter, K. von Bergmann, A. Kubetzka, and R. Wiesendanger,  Writing and deleting single magnetic skyrmions, Science {\bf 341}, 636-639 (2013).

\bibitem{McGray}
A. McGray, T. Cote, Y. Li, A. K. Petford-Long, and C. Phatak, Understanding complex magnetic spin textures with simulation-assisted Lorentz transmission electron microscopy, Physical Review Applied {\bf 15}, 044025-() (2021). 

\bibitem{BP}
A. A. Belavin and A. M. Polyakov, Metastable states of two-dimensional isotropic ferromagnets, Pis'ma Zh. Eksp.Teor. Fiz \textbf{22}, 503-506 (1975) {[}JETP Lett. \textbf{22}, 245-248 (1975){]}.

\bibitem{Manton-book} N. Manton and P. Sutcliffe, \textit{Topological Solitons}, Cambridge University Press 2004. 

\bibitem{Lectures}
E. M. Chudnovsky and J. Tejada, {\it Lectures on Magnetism}, Rinton Press (Princeton - NJ, 2006).

\bibitem{SkyrmePRC58} 
T. H. R. Skyrme, A non-linear theory of strong interactions, Proceedings of the Royal Society A \textbf{247}, 260-278 (1958).

\bibitem{Polyakov-book} A. M. Polyakov, \textit{Gauge Fields and Strings}, Harwood Academic Publishers 1987.

\bibitem{AlkStoNat01} 
U. Al Khawaja and H. T. C. Stoof, Skyrmions in a ferromagnetic Bose-Einstein condensate, Nature \textbf{411}, 918-920 (2001).

\bibitem{SonKarKivPRB93} 
S. L. Sondhi, A. Karlhede, S. A. Kivelson, and E. H. Rezayi, Skyrmions and the crossover from the integer to fractional quantum Hall effect at small Zeeman energies, Physical Review B \textbf{47}, 16419-16426 (1993).

\bibitem{StonePRB93} 
M. Stone, Magnus force on skyrmions in ferromagnets and quantum Hall systems, Physical Review B \textbf{53}, 16573-16578 (1996).

\bibitem{YeKimPRL99} 
J. Ye, Y. B. Kim, A. J. Millis, B. I. Shraiman, P. Majumdar, and Z. Tesanovic, Berry phase theory of the anomalous Hall effect: Application to colossal magnetoresistance manganites, Physical Review Letters \textbf{83}, 3737-3740 (1999).

\bibitem{WriMerRMR89} 
D. C. Wright, and N. D. Mermin, Crystalline liquids: the blue phases, Review of Modern Physics \textbf{61}, 385-433 (1989).

\bibitem{graphene}
J. Atteia, Y. Lian, and M. O. Goerbig, Skyrmion zoo in graphene at charge neutrality in a strong magnetic field, Physical Review B {\bf 103}, 035403-(29) (2021).

\bibitem{Bogdanov1989}
A. N. Bogdanov and  D. A. Yablonskii, Thermodynamically stable ``vortices'' in magnetically ordered crystals. The mixed state of magnets. Soviet
Physics JETP {\bf 68}, 101-103 (1989).

\bibitem{Bogdanov94}
A. Bogdanov and A. Hubert, Thermodynamically stable magnetic vortex states in magnetic crystals, J. Magn. Magn. Mater. {\bf 138}, 255-269 (1994).

\bibitem{Bogdanov-Nature2006}
U. K. R$\ddot{\text{o}}${\ss}ler, N. Bogdanov, and C. Pfleiderer, Spontaneous skyrmion ground states in magnetic metals, Nature \textbf{442}, 797-801 (2006).

\bibitem{Klemm}
R. A. Klemm, \textit{Layered Superconductors: Volume 1}, Oxford University Press 2011. 

\bibitem{Abanov}
A. Abanov and V. L. Pokrovsky, Skyrmion in a real film, Physical Review B {\bf 58}, R8889-R8892 (1998). 

\bibitem{CCG-PRB2012}
L. Cai, E. M. Chudnovsky, and D. A. Garanin, Collapse of skyrmions in two-dimensional ferromagnets and antiferromagnets, Physical Review B \textbf{86}, 024429-(4) (2012).

\bibitem{AFM}
A. Derras-Chouk, E. M. Chudnovsky, and D. A. Garanin, Quantum states of a skyrmion in a two-dimensional antiferromagnet, Physical Review B {\bf 103}, 224423-(8) (2021). 

\bibitem{Amel2018}
A. Derras-Chouk, E. M. Chudnovsky, and D. A. Garanin, Quantum collapse of a magnetic skyrmion, Physical Review B {\bf 98}, 024423-(9) (2018).

\bibitem{Brown}
W. F. Brown, Jr., {\it Micromagnetics} (John Wiley and Sons, New York - London, 1963).

\bibitem{MQT-book}
E. M. Chudnovsky and J. Tejada, {\it Macroscopic Quantum Tunneling of the Magnetic Moment} (Cambridge University Press, Cambridge - England, 1998).

\bibitem{Psaroudaki-PRX2017}
C. Psaroudaki, S. Hoffman, J. Klinovaja, and D. Loss, Quantum dynamics of skyrmions in chiral magnets, Physical Review X {\bf 7}, 041045-(18) (2017). 

\bibitem{Kravchuk-PRB2018}
V. P. Kravchuk, D. D. Sheka, U. K. R$\ddot{\text{o}}${\ss}ler, J. van den Brink, and Y. Gaididei, Spin eigenmodes of magnetic skyrmions and the problem of the effective skyrmion mass, Physical Review B {\bf 97}, 064403-(10) (2018). 

\bibitem{Zhang2020}
X. Zhang, Y. Zhou, K. Song, T. Park, J. Xia, M. Ezawa, X. Liu, W. Zhao, G. Zhao and S. Woo, Skyrmion-electronics: writing, deleting, reading and processing magnetic skyrmions toward spintronic applications, Journal of Physics: Condensed Matter \textbf{32}, 143001 (2020).

\end{thebibliography}
\end{document}